\documentclass[11pt,a4paper]{article}
\pdfoutput=1
\usepackage{jcappub}
\usepackage[english]{babel}
\usepackage{graphicx}
\newcommand{\gag}{g_{a\gamma}}

\begin{document}
\hfill{IPPP/16/15}

\title{Reionization during the dark ages from a cosmic axion background}

\author[a]{Carmelo Evoli,}
\author[b]{Matteo Leo,}
\author[c,d]{Alessandro Mirizzi,}
\author[e,f]{Daniele Montanino}
\affiliation[a]{Gran Sasso Science Institute, Viale Francesco Crispi 7, 67100 L'Aquila, Italy.}
\affiliation[b]{Institute for Particle Physics Phenomenology, Department of Physics, Durham University, Durham DH1 3LE, U.K.}
\affiliation[c]{Dipartimento Interateneo di Fisica ``Michelangelo Merlin,'' Via Amendola 173, 70126 Bari, Italy.}
\affiliation[d]{Istituto Nazionale di Fisica Nucleare - Sezione di Bari, Via Amendola 173, 70126 Bari, Italy.}
\affiliation[e]{Dipartimento di Matematica e Fisica ``Ennio De Giorgi,'' Via Arnesano, 73100 Lecce, Italy.}
\affiliation[f]{Istituto Nazionale di Fisica Nucleare - Sezione di Lecce, Via Arnesano, 73100 Lecce, Italy.}
\emailAdd{carmelo.evoli@gssi.infn.it, matteo.leo@durham.ac.uk, alessandro.mirizzi@ba.infn.it, daniele.montanino@le.infn.it.}

\abstract{
Recently it  has been pointed out that a cosmic background of relativistic axion-like particles (ALPs) would be produced by the primordial 
decays of heavy fields in the post-inflation epoch, contributing to the extra-radiation content in the Universe today. 
Primordial magnetic fields would trigger conversions of these ALPs into  sub-MeV photons during the dark ages.
This  photon flux would produce an early  reionization of the Universe, leaving a  significant imprint on the total optical depth to recombination $\tau$. 
Using the current measurement of  $\tau$ and the limit on the extra-radiation content $\Delta N_{\rm eff} $ by the Planck experiment we put a strong bound on the
ALP-photon conversions. 
Namely we obtain upper limits on the product of the photon-ALP coupling constant $g_{a\gamma}$ times the magnetic field strength
 $B$ down to $g_{a\gamma} B \gtrsim 6 \times 10^{-18}  \textrm{GeV}^{-1}  \textrm{nG} $ for ultralight ALPs.
}
\maketitle

\section{Introduction}

Ultralight axion-like particles (ALPs) 
 with a two-photon coupling $a \gamma \gamma$
are predicted by several extensions of the Standard Model, like four-dimensional ordinary and supersymmetric models (see
e.g.\ \cite{Coriano:2006xh,Baer:2008yd}), Kaluza-Klein theories (see e.g.\ \cite{Turok:1995ai}) and especially string theories (see e.g.\ \cite{Svrcek:2006yi,Arvanitaki:2009fg,Cicoli:2012sz}) (for a review,
see \cite{Jaeckel:2010ni}).
The two-photon vertex allows for  photon-axion mixing in external magnetic fields. This effect leading to oscillations of photons into ALPs
  is the basis of direct 
search  carried on through
\emph{(a)} Light-Shining-Through-Wall experiments (e.g.\ ALPS at DESY \cite{Ehret:2010mh}) aiming both for production and detection of ALPs in laboratory, \emph{(b)} Helioscopes
(e.g.\ CAST experiment at CERN \cite{Arik:2015cjv} and the future IAXO project \cite{Irastorza:2011gs}) aiming at detecting solar ALPs   produced by their conversions into photons inside of a strong magnet pointing towards the Sun, \emph{(c)} Haloscopes
(e.g.\ the resonant cavity ADMX experiment \cite{Asztalos:2009yp} at Livermore or the proposed
dish antenna technique \cite{Horns:2012jf}) directly searching for galactic halo dark matter ALPs  in the laboratory via their coupling to the photon.  See \cite{Graham:2015ouw} for a recent review on the experimental searches of ALPs.

The presence of cosmic magnetic fields also allows for signatures of ALPs in different astrophysical and cosmological observations.
Indeed, photon-axion conversions in large-scale cosmic magnetic fields would reduce the opacity of the universe to TeV photons, explaining the anomalous spectral hardening found in the Very-High-Energy gamma-ray 
spectra \cite{Mirizzi:2007hr,DeAngelis:2007dqd,Simet:2007sa,Mirizzi:2009aj,Horns:2012kw,Horns:2012fx,Meyer:2013pny,Meyer:2014epa,DeAngelis:2011id}. 
Moreover, in the presence of primordial magnetic fields also the Cosmic Microwave Background (CMB) spectrum would be distorted by conversions into ALPs allowing one to put
strong bounds on this mechanism \cite{Mirizzi:2005ng,Mirizzi:2006zy,Mirizzi:2009nq}. (For a wide review on axion cosmology see~\cite{Marsh:2015xka}.) Conversions of CMB photons into the magnetic fields of Galaxy Clusters have also been recently considered 
to get sharp constraints on the mixing \cite{Schlederer:2015jwa}.

An intriguing connection between ALPs and cosmology has been proposed few years ago in relation to  the possible relativistic extra-radiation in the 
Universe, parametrized as an excess with respect to the known neutrino species $\Delta N_{\rm eff}= N_{\rm eff}-3.046$ \cite{Mangano:2005cc}. The 2015 data from the 
satellite experiment Planck combined with other astrophysical measurements give $N_{\rm eff} = 3.15 \pm 0.23$ at 68 $\%$ CL \cite{Ade:2015xua}.%
\footnote{Note that the value of $N_{\rm eff}$ quoted by Planck would change depending on the dataset used and on the combination with other experiments.}
Therefore, even if the amount of possibile extra-radiation has been significantly reduced with respect to previous indications \cite{RiemerSorensen:2013ih},
the possibility of a value of $N_{\rm eff}$ larger than the standard one is still a possibility.
String theory models present opportunities to link the  extra-radiation with ALPs \cite{Cicoli:2012aq,Higaki:2012ar,Higaki:2013lra,Conlon:2013isa}. Indeed these models
 often possess heavy moduli fields which need to decay after the inflation in order not to
dominate the energy density of the Universe during the radiation era.
  In addition to the decay modes
to visible sector of Standard Model (SM) particles, these fields may also decay
to (effectively) massless weakly coupled particles from the hidden sector,
such as ALPs and hidden photons. These ultralight particles are so weakly coupled to the SM particles that do not thermalize
and contribute to the extra-radiation today \cite{Conlon:2013isa}. In particular the cosmic ALP background produced by the moduli decay 
would be present today as diffuse radiation in the energy range between 100 eV to a keV. At this regard
it has been proposed  that the  observed soft X-ray excesses in many Galaxy Clusters may be explained by the conversion of the
 cosmic ALP background radiation into photons in the Cluster magnetic field \cite{Conlon:2013txa,Angus:2013sua,Powell:2014mda}. 
 
 Assuming  that the  amount of extra-radiation compatible with the latest Planck data is    composed  by ALPs produced by  moduli decays, 
this would have a strong impact on the thermal history of the Universe. 
 Indeed the presence of a primordial magnetic field would inevitably
trigger 
conversions between these cosmic ALPs and photons, whose impact on cosmological observables would be useful to get strong constraints 
on the photon-ALP coupling as discussed in \cite{Higaki:2013qka}. In this context, we focus on ALP-photon conversions during the dark
ages (at redshift $6 \lesssim z \lesssim 1100$). The produced high-energy photons [$E \lesssim {\mathcal O}$(MeV)]  would ionize the recently formed light atoms of Hydrogen and Helium~\cite{Mapelli:2006}. 
Remarkably,  even conversion probabilities at a level 
$\lesssim 10^{-9}$   would have 
a dramatic impact on the optical depth of the Universe $\tau$. Using the latest measurement of $\tau$ by the Planck experiment  \cite{Ade:2015xua} it would be 
possible to put strong constraints on this mechanism. 
Our work is devoted to discuss in detail this effect.
In Sec.~2 we characterize the cosmic ALP background flux produced by moduli decays. In Sec.~3   we discuss the conversions of these cosmic ALPs into photons in the primordial $B$-fields. 
In Sec.~4  we show the impact of the ALP-photon conversions on the reionization and we present our  bound from the cosmic optical depth. Finally in Sec.~5 we comment on our results and we conclude. 
It follows an Appendix where we present technical details on the derivation of the approximate expression of the ALP-photon conversion probability used in our work. 

\section{Cosmic ALP background}

A  generic feature of the four-dimensional effective theories arising from compactifications of string
theory is the presence of massive scalar particles, dubbed moduli, with gravitational strength coupling.
The total decay rate of moduli (into ALPs and SM particles) during  the post-inflation epoch
is given by  \cite{Conlon:2013isa}
\begin{equation}
\Gamma_\Phi = \frac{\kappa^2}{4 \pi}\frac{m_{\Phi}^3}{M_{\rm pl}^2} \,\ , 
\end{equation}
where $M_{\rm pl}= 2.4 \times 10^{18}$~GeV is the Planck mass, $m_{\Phi}$ is the modulus mass and $\kappa$ is a constant of order ${\mathcal O}(1)$. The SM particles from moduli decays would rapidly thermalize at the reheating temperature
\begin{equation}
T_{\rm reheat} \sim \frac{m_{\Phi}^{3/2}}{M_{\rm pl}^{1/2}}  \sim 1\,\ \textrm{GeV} \left(\frac{m_\Phi}{10^6 \,\ \textrm{GeV}} \right)^{3/2} \,\ .\label{eq:reheat}
\end{equation}
Moduli can also decay into
light states from the hidden sector, like ALPs, with initial energy $\varepsilon_\textrm{a} = m_{\Phi}/2$ and
 decay rate  given by   $\Gamma_\textrm{a}=B_\textrm{a} \Gamma_\Phi$ with $B_a$ the  branching ratio in ALPs.
If these ALPs are weakly coupled to SM, they do not thermalize
and remain till now as dark radiation, with typical energy today $\varepsilon_\textrm{a} \sim {\mathcal O}(100)$ eV
for moduli masses $m_{\Phi} \sim 10^6$  GeV. 
The ALP spectrum has been calculated as in \cite{Conlon:2013isa}. The Boltzmann equations for moduli number density $ n_\Phi$,  and  for ALPs and standard
 radiation
$\rho$  are
respectively
\begin{eqnarray}
\dot n_\Phi+3Hn_\Phi&=&-\Gamma_\Phi n_\Phi(t)\, \, ,\nonumber\\
\dot\rho+4H\rho&=&\Gamma_\Phi m_\Phi n_\Phi(t)\, \, .
\label{eq:densev}\end{eqnarray}
The Friedmann-Robertson-Walker equation 
\begin{equation}
H(t)=\frac{\dot R}{R}=\sqrt{\frac{m_\Phi n_\Phi(t)+\rho(t)}{3M_{\rm pl}^2}}\, \, ,
\end{equation}
closes the system, where $H(t)$ is the Hubble parameter, expressed in terms of the scale factor of the Universe $R(t)$.
We assign $R(t_0)=1$ with $t_0$ the age of the Universe. 
With  the following change of  variables $t=\theta\cdot\Gamma_\Phi^{-1}$, $n_\Phi(t)=\nu(\theta)\cdot N_\Phi\eta^3$, $\rho(t)=\sigma(\theta)\cdot N_\Phi m_\Phi\eta^3$, $R(t)= x(\theta)/\eta$, $H(t)=x^\prime(\theta)/x(\theta)\cdot \Gamma_\Phi$, with $N_\Phi$ initial comoving number of moduli and
\begin{equation}
\eta=\left(\frac{3M_{\rm pl}^2}{N_\Phi m_\Phi}\right)^{1/3}\Gamma_\Phi^{2/3}\,\, ,
\label{eq:zeta}\end{equation}
with initial conditions $x(0)=0$, $\nu x^3\to1$ and $\sigma x^4\to 0$ for $\theta\to 0$, the previous equations can be reduced to the following {\em universal} (i.e., independent from physical quantities) equation
\begin{equation}
h(\theta)=\frac{x^\prime(\theta)}{x(\theta)}=\left[\frac{\text{e}^{-\theta}}{x^3(\theta)}+\frac{1}{x^4(\theta)}\int_0^\theta d\xi\, \text{e}^{-\xi}\,x(\xi)\right]^{1/2}\,\, .
\label{eq:evol1}
\end{equation}
This equation can be integrated numerically. For small $\theta$ the evolution is matter dominated ($x\sim \theta^{2/3}$) while for large $\theta$ the evolution is radiation dominated ($x\sim \theta^{1/2}$).

The number density of ALPs per unit of energy and comoving volume $\mathcal{N}_\textrm{a}(E,t)$ follows the Boltzmann equation
\begin{equation}
\left[\frac{\partial}{\partial t} -H(t)E\frac{\partial}{\partial E}-H(t)\right] \mathcal{N}_\textrm{a}=2 B_\textrm{a}\Gamma_\Phi\delta\left(E-\frac{m_\Phi}{2}\right)N_\Phi \text{e}^{-\Gamma_\Phi t}
 \,\, ,
\label{eq:Boltz1}\end{equation}
where we suppose that the moduli decay at rest (since $T_{\rm reheat}\ll m_\Phi$) and the factor $2$ on the right side accounts for the fact that two ALPs are produced for each moduli decay. Using the comoving energy  $E_0=E R(t)=Ex/\eta$ and $x$ as independent variables we can rewrite the previous equation as
\begin{equation}
\left(x\frac{\partial}{\partial x}-1\right) \mathcal{N}_\textrm{a}=\frac{2B_\textrm{a}}{h(\theta(x))}\frac{x^2}{\eta E_0}\delta\left(x-\eta\frac{2E_0}{m_\Phi}\right)N_\Phi \text{e}^{-\theta(x)}\,\, .
\end{equation}
This equation can be integrated from $x=0$ to $x> 2\eta E_0/m_\Phi$ giving
\begin{equation}
\frac{dn_\textrm{a}(E,z)}{dE}=(1+z)^3 \mathcal{N}_\textrm{a}(E,t)=\frac{2N_\Phi B_\textrm{a}}{\varepsilon_\textrm{a}}(1+z)^2\varphi\left(\frac{E}{(1+z)\varepsilon_\textrm{a}}\right)\,\, ,
\label{eq:dna}\end{equation}
where
\begin{equation}
\varphi(x)=-\frac{d}{dx}\text{e}^{-\theta(x)}\,\, ,
\end{equation}
and $\varepsilon_\textrm{a}=m_\Phi/2\eta$ and we have used $x/\eta=R(t)=(1+z)^{-1}$ and $h=(x\theta^\prime(x))^{-1}$.
The factor $n_\textrm{a}=2N_\Phi B_\textrm{a}$ is the total density of ALPs today. The function $\varphi(x)$ can be calculated numerically from Eq.~(\ref{eq:evol1}).
A handy and very accurate approximation is a quasi-thermal spectrum given by
\begin{equation}
\varphi(x)\simeq\frac{\alpha K^{\alpha/\beta}}{\Gamma\left(\frac{\alpha}{\beta}\right)}
x^{\alpha}\textrm{exp}(-K x^{\beta}) \,\ ,
\end{equation}
where $\Gamma$ is the Euler gamma function, $\alpha=1.434$, $\beta=1.980$ and $K=0.445$.

Since $N_\Phi$ and $B_\textrm{a}$ are unknown, it is convenient to write Eq.~(\ref{eq:dna}) in terms of observable parameters.  The density of SM radiation after moduli decay can be written as 
\begin{equation}
\rho_{\textrm{SM}}(t_D)=\eta^{-1} J\cdot m_\Phi N_\Phi(1-B_\textrm{a})\cdot\frac{1}{R^4(t_D)}\,\, ,
\label{eq:rhoSM}
\end{equation}
where $t_D\sim {\mathcal O}(10\, \Gamma_\Phi^{-1})$ and $J=\int_0^\infty d\theta\, \text{e}^{-\theta}x(\theta)\simeq1.078$. This radiation can be observed nowadays as 
CMB radiation. Using $\rho\propto (g^*)^{-1/3}R^{-4}$ we have
\begin{equation}
\rho_{\textrm{CMB}}=\left(\frac{11}{4}\right)^{1/3}\cdot\left(\frac{22}{43}\right)\cdot
\left[\frac{g^*(T_D)}{g^*(T_\nu)}\right]^{1/3}\cdot R^4(t_D)\cdot\rho_{\textrm{SM}}(t_D)\,\, ,
\label{eq:rhoCMB}\end{equation}
where $\rho_{\textrm{CMB}}=2\times10^{-51}$~GeV$^4$ is the present CMB energy density, $g^*(T_\nu)=10.25$ the number of degree of freedom at the neutrino decoupling, $g^*(T_D)$ the number of degree of freedom at the end of moduli decay, the factor $(11/4)^{1/3}$ accounts for the photon reheating due to $\text{e}^+\text{e}^-$ annihilation, while $22/43$ is the fraction of radiation in visible form after neutrino decoupling.

Note that in principle is not easy to determine $g^\ast(T_D)$ because $T_D\lesssim T_\textrm{reheat}$ is not a well defined quantity. However, is reasonable to expect that moduli decay occurs well before QCD phase transition [Eq.~(\ref{eq:reheat})], where $g^*(T_\textrm{QCD})=61.75$. For $T_D>T_\textrm{QCD}$, $g^*(T)$ is a smooth function of temperature. We will see that the final ALPs density and energy depends mildly from $g^*(T_D)$ (as [$g^*(T_D)]^{1/12}$). Moreover,  $g^*(T_D)$ can be reabsorbed by a redefinition the (unknown) moduli mass. For this reason the exact determination of $g^*(T_D)$ is not crucial.

We also remind that $B_\textrm{a}$ will depend to the final observable extra-radiation  $\Delta N_{\rm eff}$ by the relation \cite{Conlon:2013isa}:
\begin{equation}
\Delta N_{\rm eff}=\frac{43}{7}\frac{B_\textrm{a}}{1-B_\textrm{a}}\left[\frac{g^*(T_\nu)}{g^*(T_D)}\right]^{1/3}\,\, ,
\end{equation}
so that 
\begin{equation}
B_\textrm{a}\simeq 0.29\left(\frac{g^*_D}{61.75}\right)^{1/3}\Delta N_{\rm eff}\,\, .
\label{eq:branch}
\end{equation}
Eq.~(\ref{eq:rhoCMB}) can be inverted to obtain $N_\Phi$. After straightforward calculations we can infer the total density of ALPs at $t=t_0$
\begin{equation}
n_\textrm{a}=2N_\Phi B_\textrm{a}=133\,{\rm m}^{-3}\left(\frac{\widetilde{m}_\Phi}{\rm PeV}\right)^{1/2}\Delta N_{\rm eff} \,\, ,
\label{eq:ALPdens}\end{equation}
where we have defined the ``effective'' moduli mass
\begin{equation}
\widetilde{m}_\Phi=(1-B_\textrm{a})^{1/2}\kappa^2\left(\frac{g^*_D}{61.75}\right)^{1/6} m_\Phi\,\, .
\label{eq:effmass}\end{equation}

\begin{figure}
\centering
\includegraphics[height=10cm, width=10cm]{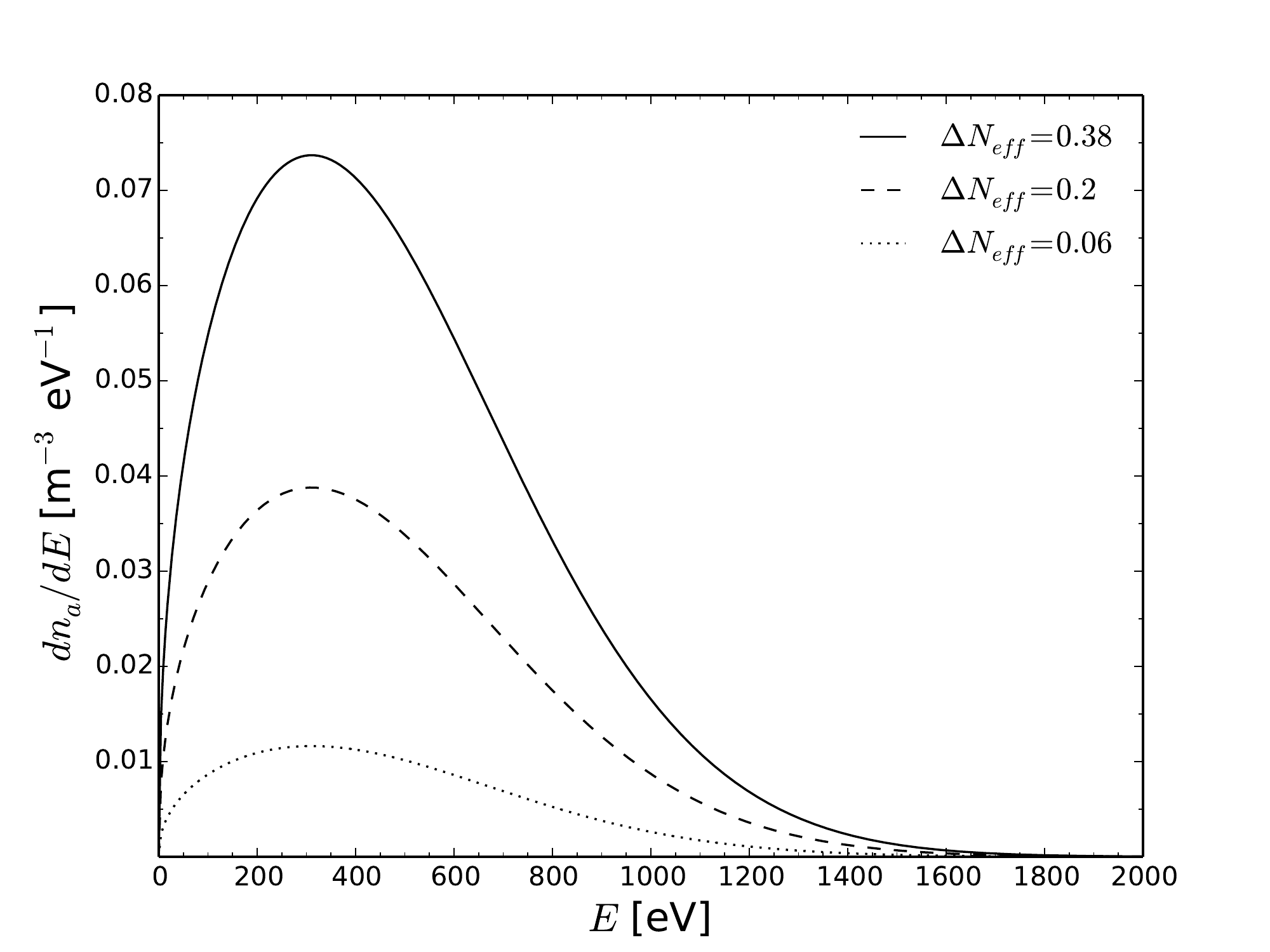}
\caption{Cosmic ALP spectrum today for different values of $\Delta N_{\rm eff}$. (see the text for details)
\label{cab}}
\end{figure}

The actual ($z=0$) ALP spectrum thus can be written as
\begin{equation}
\frac{d n_\textrm{a} (E)}{d E}
= \frac{n_\textrm{a}}{\varepsilon_\textrm{a}}
\varphi\left(\frac{E}{\varepsilon_\textrm{a}} \right) \,\, ,
\label{eq:dna1}
\end{equation}
with 
\begin{equation}
\varepsilon_\textrm{a}=412\left(\frac{\widetilde{m}_\Phi}{\rm PeV}\right)^{-1/2}\,\textrm{eV}\,\, .
\end{equation}
The ALP spectrum in Eq.~(\ref{eq:dna1}) is 
represented in Fig.~\ref{cab} for the benchmark parameters in parenthesis of Eq.~(\ref{eq:branch})--(\ref{eq:effmass}).
For the  amount of ALP extra-radiation we assume $\Delta N_{\rm eff}=0.38$ (continuous curve) corresponding to the $1\sigma$ range of Planck 2015 
analysis \cite{Ade:2015xua}, $\Delta N_{\rm eff}=0.2$ (dashed curve) and $\Delta N_{\rm eff}=0.06$ (dotted curve) corresponding to the  $2\sigma$  sensitivity
of the future EUCLID mission \cite{Basse:2013zua}.

\section{ALP-photon conversions in the Early Universe}

\subsection{ALP-photon conversions}

Photon-ALP mixing occurs in the presence of an external magnetic field ${\bf B}$ due to the Lagrangian~\cite{Raffelt:1987im,sikivie,Anselm:1987vj}
\begin{equation}
\label{mr}
{\cal L}_{a\gamma}=-\frac{1}{4} \,\gag
F_{\mu\nu}\tilde{F}^{\mu\nu}a=\gag \, {\bf E}\cdot{\bf B}\,a~,
\end{equation}
where $\gag$ is the photon-ALP coupling constant (which has the dimension of an inverse energy).

We  assume a beam
of energy $E$ propagating
along the $x_3$ direction in a cold ionized and magnetized medium.
At first we restrict our attention to the case in which the magnetic field ${\bf B}$ is homogeneous. We denote by ${\bf B}_T$ the transverse magnetic field, namely its component in the plane normal to the beam direction and we choose the $x_2$-axis along ${\bf B}_T$ so that $B_1$ vanishes. The linear photon polarization state parallel to the transverse field direction ${\bf B}_T$ is then denoted by $A_{\parallel}$ and the orthogonal one by $A_{\perp}$.
 Neglecting for the moment incoherent scattering,  the linearized equations of motion for the ALP state $a$  and for the two polarization states of the photons are~\cite{Raffelt:1987im}
\begin{equation}
\label{vne}
\left(E- i \frac{\partial}{\partial x_3} + {\mathcal M} \right)
 \left(\begin{array}{c} A_{\perp} \\ A_{\parallel} \\a
 \end{array}\right)
= 0 \,\ ,
\end{equation}
where the mixing matrix can be written as~\cite{Mirizzi:2005ng,Mirizzi:2006zy}
\begin{equation}
{\cal M}_0 = \left(\begin{array}{ccc}
\Delta_{ \perp}   & 0 & 0 \\
0 &  \Delta_{ \parallel}   & \Delta_{a \gamma}  \\
0 & \Delta_{a \gamma} & \Delta_\textrm{a} 
\end{array}\right)~,
\label{eq:massgen}
\end{equation}
whose elements are~\cite{Raffelt:1987im} $\Delta_\perp \equiv \Delta_{\rm pl} + \Delta_{\perp}^{\rm CM},$ $ \Delta_\parallel \equiv \Delta_{\rm pl} + \Delta_{\parallel}^{\rm CM} + \Delta_{\textrm{CMB}},$ $\Delta_{a\gamma} \equiv {g_{a\gamma} B_T}/{2} $ and $\Delta_\textrm{a} \equiv - {m_\textrm{a}^2}/{2E}$, where $m_\textrm{a}$ is the ALP mass. The term $\Delta_{\rm pl} \equiv -{\omega^2_{\rm pl}}/{2E}$ accounts for plasma effects, where $\omega_{\rm pl}$ is the plasma frequency expressed as a function of the electron density in the medium $n_e$ as $\omega_{\rm pl} \simeq 3.69 \times 10^{- 11} \sqrt{n_e /{\rm cm}^{- 3}} \, {\rm eV}$. The  terms $\Delta_{\parallel,\perp}^{\rm CM}$ describe the Cotton-Mouton  effect, i.e.~the birefringence of fluids in the presence of a transverse magnetic field.  A vacuum Cotton-Mouton effect is expected from QED one-loop corrections to the photon polarization in the presence of an external magnetic field $\Delta_\mathrm{QED} = |\Delta_{\perp}^{\rm CM}- \Delta_{\parallel}^{\rm CM}| \propto B^2_T$, but this effect is completely negligible in the present context, as we will show in Sec.~3.4. 
Recently it has been realized that also background photons can contribute to the  photon polarization. At this regard a guaranteed
contribution is provided by the CMB radiation, leading to $\Delta_{\textrm{CMB}} \propto \rho_{\textrm{CMB}}$ \cite{Dobrynina:2014qba}. We will show in  Sec.~3.4 how this term would play a crucial role
for the development of the conversions at large redshift during the recombination epoch.
 An off-diagonal $\Delta_{R}$ would induce the Faraday rotation, which is however totally irrelevant at the energies of our interest, and so it has been dropped. For the relevant parameters,  we numerically find 
\begin{eqnarray}  
\Delta_{a\gamma}&\simeq &   1.52\times10^{-8} \left(\frac{g_{a\gamma}}{10^{-17}\textrm{GeV}^{-1}} \right)
\left(\frac{B_T}{10^{-9}\,\rm G}\right) {\rm Mpc}^{-1}
\nonumber\,,\\  
\Delta_\textrm{a} &\simeq &
 -7.8 \times 10^{5} \left(\frac{m_\textrm{a}}{10^{-10} 
        {\rm eV}}\right)^2 \left(\frac{E}{{\rm keV}} \right)^{-1} {\rm Mpc}^{-1}
\nonumber\,,\\  
\Delta_{\rm pl}&\simeq & 
  -1.1\times10^{-2}\left(\frac{E}{{\rm keV}}\right)^{-1}
         \left(\frac{n_e}{10^{-7} \,{\rm cm}^{-3}}\right) {\rm Mpc}^{-1}
\nonumber\,,\\
\Delta_{\rm QED}&\simeq & 
4.1\times10^{-18}\left(\frac{E}{{\rm keV}}\right)
\left(\frac{B_T}{10^{-9}\,\rm G}\right)^2 {\rm Mpc}^{-1} \nonumber\,,\\
\Delta_{\textrm{CMB}}&\simeq & 0.80 \times 10^{-10} \left(\frac{E}{{\rm keV}} \right)  {\rm Mpc}^{-1} \,\ .
\label{eq:Delta0}\end{eqnarray}

For the above estimates that we will use in the following as benchmark values, we refer to the following physical inputs: The strength of $B$-fields and the electron density $n_e$ in the previous equations are typical values for the intergalactic medium.
Moreover, we note that in the equations enter only the product $g_{a\gamma} B_T$. Since we will be interested in possible ALP-photon conversions during the recombination epoch in the Early Universe,
we require $g_{a\gamma} B_T \lesssim 10^{-13}$~GeV~nG for $m_\textrm{a} \lesssim 10^{-9}$~eV in order to avoid an excessive distortion of the CMB spectrum \cite{Mirizzi:2009nq}  caused by the
conversions during that phase. 
This requirement satisfies also the direct experimental bound on $g_{a \gamma}$ obtained by the CAST experiment \cite{Arik:2015cjv}, the  globular-cluster limit \cite{Ayala:2014pea}
and the one  for ultra-light ALPs from the absence of $\gamma$-rays from SN~1987A \cite{Payez:2014xsa}.
Concerning the ALP energy today the bulk of the spectrum is in the range $E\in [100, 800]$~eV (see Fig.~1).
Finally, we will assume $m_\textrm{a} \ll \omega_{\rm pl}$, neglecting the term $\Delta_\textrm{a}$ in the evolution, since a  value of $m_a$ much larger than
$ \omega_{\rm pl}$
would suppress the conversions. 

We note that in the case of a homogeneous magnetic field, the ALP-photon mixing described by Eq.~(\ref{vne}) reduces to a $2\times2$ problem involving only
the $(A_{\parallel}, a)$.
Since in the following we will consider  cosmic ALP inter-converting into photons in the primordial magnetic fields, we have to deal with a more general situation than the one of the homogeneous case considered above. 
The possible existence of a primordial magnetic field of cosmological  origin has been the
subject of an intense investigation during the last few decades~\cite{Durrer:2013pga}.  However despite these efforts,
there is 
no astrophysical evidence  for primordial  magnetic fields, and only upper limits are reported.
In particular primordial $B$-fields would have an impact on  CMB anisotropies. 
Overall, Planck data constrain the amplitude of
primordial $B$-fields to be less than a few nG coherent on a scale $l \sim {\mathcal O}$(1 Mpc) \cite{Ade:2015cva}.
Therefore, the primordial $B$-field  has a domain-like structure with size set by its coherence length.  The strength of ${\bf B}$ is assumed to be the same in every domain, but  its direction changes in a random way  from one cell to another. 
Because of this, $A_{\parallel}$ in one region or cell is not the same as $A_{\parallel}$ in the next one. 
Therefore the propagation over many magnetic domains represents  a full 3-dimensional case.
In order to treat this problem   we consider the $x_1$, $x_2$, $x_3$ coordinate system
and expand the photon polarization states on this basis of coordinates, i.e. $(A_1, A_2)$.

We take the coordinate basis as fixed once and for all, and -- labelling with $\psi_p$ the angle between $B_T$ and the $x_2$ axis in the generic $p$-th domain ($1 \leq p \leq n$) -- we treat every $\psi_p$  as a random variable in the range $0 \leq \psi_p < 2 \pi$. During their propagation with a total path $L$, the beam crosses $n = L/l$ domains, where $l$ is the size of each domain. Then the set $ \{(B_T)_p \}_{1 \leq p \leq n}$ represents a given random realization of the beam propagation corresponding to the the angles
 $ \{\psi_p \}_{1 \leq p \leq n}$. Since we are interested only in the average of the conversion probability, we do not need to make any assumption on the probability distribution function of the $(B_T)_p$'s.  In the following for simplicity we denote with $B$  and $B_T$ the r.m.s.\ of the magnetic field and of its transversal component on all possible configurations respectively ($B\equiv\sqrt{\langle B^2\rangle}$). For symmetry reasons of course we have $B_T^2=2B^2/3$. We remark that this model with constant magnetic field in each cell is unrealistic because the condition $\nabla\cdot\mathbf{B}=0$ cannot be satisfied on the boundary of cells, however is a good approximation for practical purposes. In the Appendix we will show briefly how to relax this approximation. 

 Accordingly, in each domain the matrix ${\cal M}$ takes the form~\cite{Mirizzi:2005ng}
 \begin{equation}
\label{aa8MR}
{\cal M}_p = \left(
\begin{array}{ccc}
\Delta_{xx} & \Delta_{xy} & \Delta_{a\gamma} \, \sin\psi_p\\
\Delta_{yx} & \Delta_{yy} & \Delta_{a\gamma} \, \cos\psi_p\\
\Delta_{a\gamma} \, \sin\psi_p& \Delta_{a\gamma} \, \cos\psi_p& \Delta_{a} \\
\end{array}
\right)~,
\end{equation} 
with 
\begin{equation}
\Delta_{xx} = \Delta_\parallel \, \sin^2 \psi_p + \Delta_\perp \cos^2 \psi_p~,
\end{equation}
\begin{equation}
\Delta_{xy} = \Delta_{yx}=(\Delta_\parallel -\Delta_\perp) \sin\psi_p \, \cos\psi_p~,
\end{equation}
\begin{equation}
\Delta_{yy} = \Delta_\parallel \cos^2 \psi_p + \Delta_\perp \sin^2 \psi_p~.
\end{equation}
In general the solution of this problem can be obtained in terms of the product of transfer functions for the ALP-photon ensemble across the different 
magnetic domains~\cite{Bassan:2010ya}. However in the case we are dealing with, since we expect small conversion probabilities, 
we will show in the Appendix that is possible to adopt a perturbative approach for this calculation.

\subsection{Universe expansion}
The evolution of the ALP-photon beam in the homogeneous and isotropic Universe can be characterized considering 
the time $t \simeq x_3$ as evolution variable. 
Furthermore,  one has to take into account
the expansion of the Universe \cite{Christensson:2002ig}.
This can be accounted expressing the evolution in terms of the redshift by means of 
\begin{equation}
\frac{dt}{dz}= -\frac{1}{(1+z) H_0 \sqrt{\Omega_{\Lambda} +
\Omega_{\rm m} (1+z)^3} }\,\ ,
\end{equation}
where according to the latest Planck data, the Hubble constant parameter
$H_0= (67.8 \pm 0.9)$ km s$^{-1}$ Mpc$^{-1}$, the matter density parameter
$\Omega_{\rm m}= 0.308 \pm 0.012$ and the  dark energy density
$\Omega_\Lambda = 0.6911\pm 0.0062$  \cite{Ade:2015xua}. 

The   redshift effect leaves   unaltered the equations
of motion [Eq.~(\ref{vne})], after an appropriate rescaling
of the different parameters.
Since the number density of the electrons traces that of matter, and the average number density of electrons goes as the third power of the size of
the Universe, we obtain the relationship
\begin{equation}
n_e(z)= n_{e,0} (1+z)^3 \,\ .
\end{equation}
Note that the thermal history of the electron density can be rather complicated due to the presence of
reionization (at $z \lesssim 30$) and recombination into H and He (at $z \lesssim 1100$) \cite{Mirizzi:2009nq}. However, since in our study we will consider
ALP energies much larger than the binding  energies of these light atoms, we will always consider the contribution of $n_e$ to the plasma 
density $\omega_{\rm pl}$ as if the electrons were all free, with a density obtained simply redshifting their actual one.  

\begin{figure}
\centering
\includegraphics[height=18cm, width=18cm]{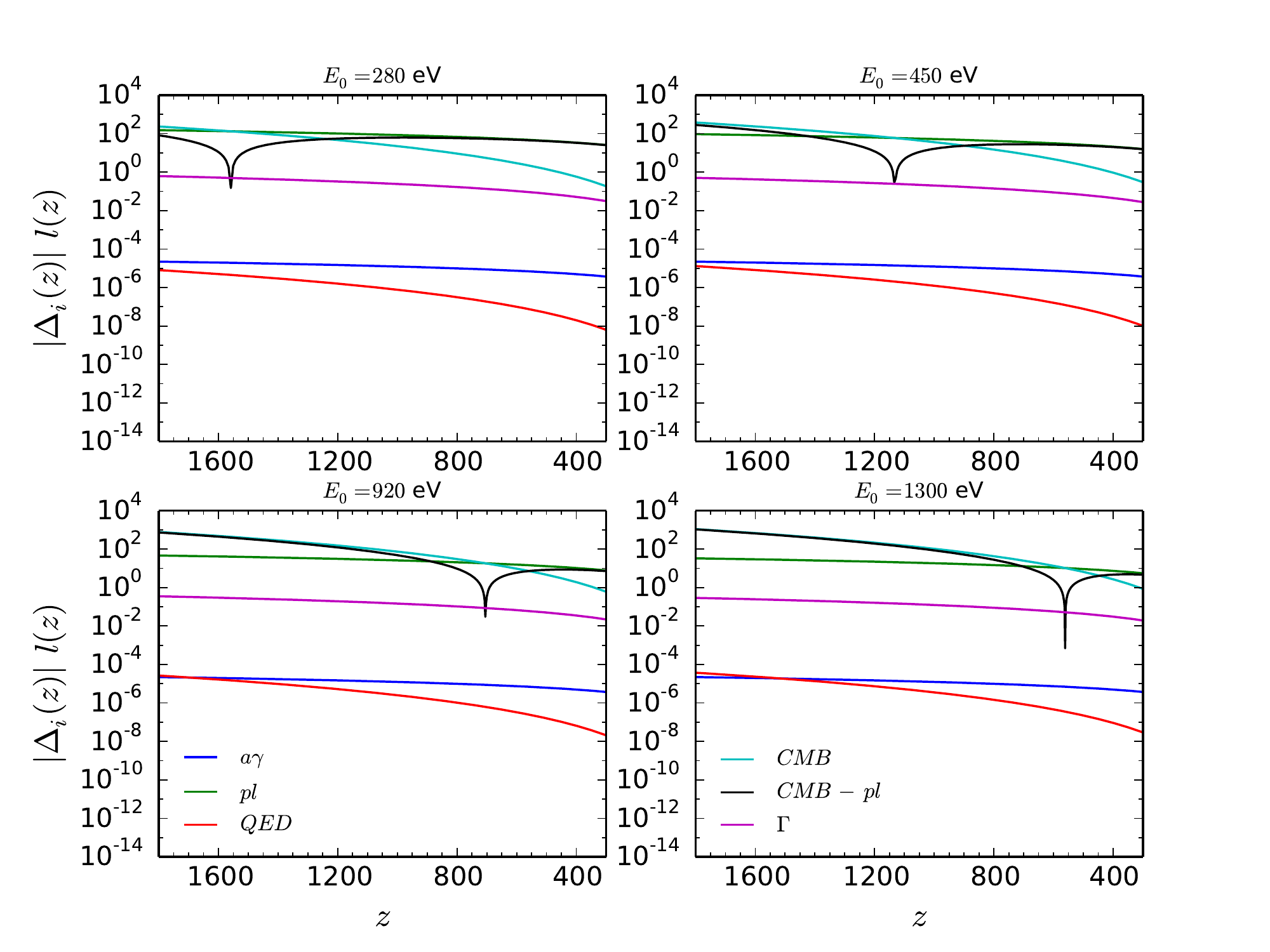}
\caption{Evolution with redshift of the different 
quantities $\Delta_{a\gamma}$, $\Delta_{\rm pl}$, $\Delta_{\rm QED}$,  
$|\Delta_{\rm CMB}+\Delta_{\rm pl}|$   and $\Gamma$ multiplied for $l(z)$. 
\label{param}}
\end{figure}

For the magnetic field, being frozen into the medium we have the relation
\begin{equation}
B(z)= (1+z)^2 B_0 \,\ ,
\end{equation}
where the size of each cell scales as
\begin{equation}
l(z)=\frac{l_0}{(1+z)} \,\, ,
\end{equation}
while the energy of the beam scales as
\begin{equation}
E(z)=E_0(1+z) \,\, ,
\end{equation}
where with subscript $0$ we indicate the today values of the different quantities.
Considering this redshift relation we find that the quantities in Eq.~(\ref{eq:Delta0}) evolve as
\begin{eqnarray}  
\Delta_{a\gamma}&=&  {\Delta^{0}_{a\gamma}} (1+z)^2
\nonumber\,,\\  
\Delta_\textrm{a} &=& \frac{\Delta^0_\textrm{a}}{(1+z)}
\nonumber\,,\\  
\Delta_{\rm pl}&=& 
\Delta^0_{\rm pl} (1+z)^2
\nonumber\,,\\
\Delta_{\rm QED}&=& \Delta^0_{\rm QED} (1+z)^5 \nonumber\,,\\
\Delta_{\rm CMB}&=& \Delta^0_{\rm CMB} (1+z)^5
\, ,
\label{eq:Deltaz}
\end{eqnarray}
where the supescript $0$ indicates the today value.
In Fig.~\ref{param} we show the evolution of the different quantities of  Eq.~(\ref{eq:Deltaz}) [multiplied by $l(z)$] in function of the redshift $z$ at different representative
energies of the ALP spectrum for  $g_{a\gamma} B=10^{-17}  \textrm{GeV}^{-1}  \textrm{nG}$.  
From the scaling law  one realizes that for $z \gg 10^3$, $\Delta_{\rm CMB} $ exceeds
all the other quantities. In particular, being larger than the off-diagonal quantity  $\Delta_{a \gamma}$, it suppresses the ALP oscillations. Therefore, in the following we will start the characterization of the conversions at $z \lesssim 1800$.  [Note that CMB effect was not included in \cite{Higaki:2013qka}, where effects of conversions at earlier epochs (e.g. during Big Bang Nucleosynthesis) was discussed.]
In this range the $\Delta_{\rm QED}$ term is completely negligible.
As the Universe expands eventually there is a point where $|\Delta_{\rm CMB}+\Delta_{\rm pl}|$ cancels (continuous black curve).
When this occurs the diagonal term in the Hamiltonian [Eq.~(\ref{eq:Delta0})]  vanishes. This implies that one can have large \emph{resonant} conversion effects there.
At lower redshift one expects the ALP conversions to be suppressed again due the dominance of the plasma term $\Delta_{\rm pl}$ over $\Delta_{a\gamma}$.
Therefore, we expect only a narrow range in $z$ where ALP-photon conversions would occur.

We mention that in the presence of a finite ALP mass $(m_{\rm a} \lesssim 10^{-10}$~eV) it would be possible to encounter also a $\omega_{\rm pl}=m_{\rm a}$ resonance that would leave a further imprint on the pattern of conversions. Conversely, for an ALP mass greater than $10^{-10}$~eV no resonance is allowed and then $\gamma\leftrightarrow a$ oscillations would be suppressed.

\subsection{Photon absorption during recombination}

\begin{figure}
\centering
\includegraphics[height=12cm, width=12cm]{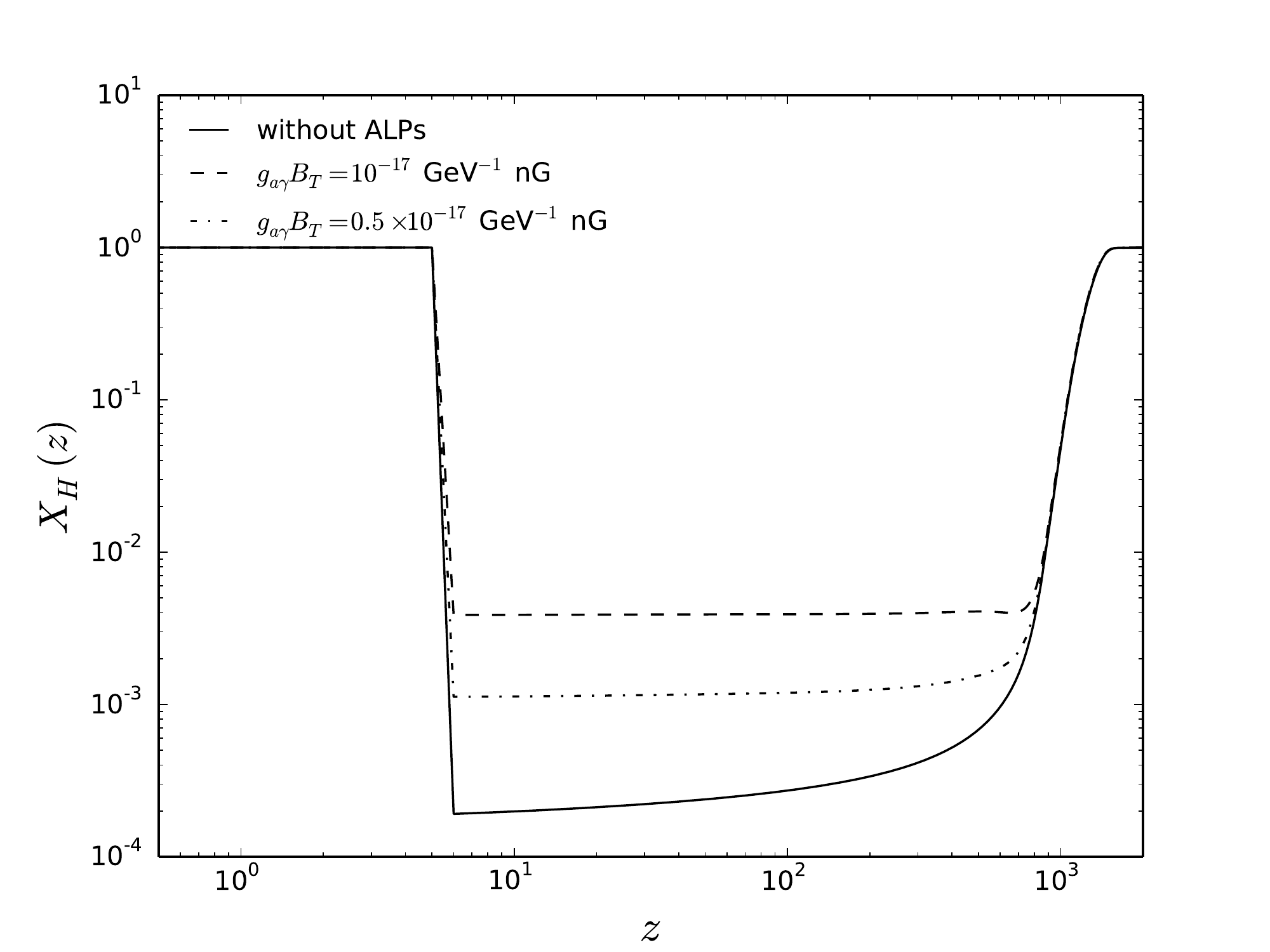}
\caption{Fraction of ionized hydrogen $X_{\rm H}$ (normalized to 1 for full ionization) in function of redshift $z$ without (continuous curve) and with ALP conversions 
for $g_{a\gamma} B=10^{-17}  \textrm{GeV}^{-1}  \textrm{nG}$ (dashed curve) and $g_{a\gamma} B=5 \times 10^{-18}  \textrm{GeV}^{-1} \textrm{nG}$  (dot-dashed curve).
We assume $\Delta N_{\rm eff}=0.38$. 
\label{XH}}
\end{figure}

As discussed before, ALP conversions into photons start 
at $z \sim 10^3$, i.e.~in the phase of matter-radiation decoupling started at $z \sim 1100$. 
At this epoch, at $T \sim 0.3$ eV, the process $H + \gamma \leftrightarrow p + \text{e}^-$ goes out of equilibrium, 
leading the the \emph{recombination} of electrons and protons into Hydrogen and Helium.
Following this event, in the so-called \emph{dark ages}, most of the intergalactic matter was composed largely of neutral Hydrogen and Helium, being 
the fraction of ionized Hydrogen (normalized to 1 for full ionization)
$X_{\rm H} \sim 10^{-4}$. This epoch lasted from $z \sim 1100$ to $z \sim 30$.
The emergence of the first luminous sources at $z \sim 30$ corresponds to the period during which the gas in the Universe went from being almost completely neutral to a state in which it became almost completely ionized.
In fact, the \emph{reionization} of the Universe was complete at about redshift $z \sim 6$~\cite{Fan:2005eq}.
The reionization sources may have been stars, galaxies, quasars, or some combination of the above~\cite{Ferrara:2014sda}. 
A large amount of experimental efforts, as the LOFAR telescope~\cite{Jensen:2013fha}, aims at  measuring the neutral gas fraction in the Universe as a function of redshift in order to have a more detailed understanding of the astrophysical processes at these earlier epochs.

In Fig.~\ref{XH} we show as a continuos curve the evolution of  $X_{\rm H}$ after recombination in the standard scenario (as computed with the RECFAST code~\cite{RECFAST}), and assuming that reionization has occurred instantaneously at $z = 6$.
Photons produced by ALP conversions during the dark ages have an energy $E \in [0.1, 0.8]$ MeV. Therefore, they
are much more energetic than the CMB photons and would ionize the recently formed neutral atoms much earlier than the standard reionization.

The photon absorption on electrons and atoms adds a damping term in the ALP-photon Hamiltonian \cite{Mirizzi:2009aj}
\begin{equation}
{\mathcal M} \to {\mathcal M} + \textrm{diag}(- i \frac{\Gamma}{2}, -i \frac{\Gamma}{2}, 0) \,\ .
\end{equation}
In the absorption term
one has to take into account the photo-electric and to the Compton effect.
The absorption term assumes then the form
\begin{equation}
\Gamma = \sigma_{\rm KN} n_e + \sigma^{\rm PE}_{\rm H} n_{\rm H} + \sigma^{\rm PE}_{\rm He} n_{\rm He} \,\ ,
\end{equation}
where $\sigma_{\rm KN}$ refers to the Klein-Nishina cross-sections for the Compton effect on (free or bound) electrons, 
while $\sigma^{\rm PE}_{\rm H} , \sigma^{\rm PE}_{\rm He}$ refer to the photo-electric effect on H and He, respectively. 
These latter cross sections are taken from~\cite{Verner:1996th}.%
\footnote{Fortran subroutines for the calculation of photo-electric cross sections can be found at the following URL:  http://www.pa.uky.edu/$\sim$verner/photo.html}
Moreover we take
\begin{eqnarray}
n_e &=& \left(1- \frac{Y_{\rm He}}{2} \right) n_B \,\ , \\
n_{\rm H} &=& \left(1- {Y_{\rm He}} \right) n_B \,\ , \\
n_{\rm He} &=& \frac{Y_{\rm He}}{4} n_B \,\ .
\end{eqnarray}
where ${Y_{\rm He}}=0.2465$ is the Helium fraction and $n_B= 2.482 \times 10^{-7}$ cm$^{-3}$  is the baryon density~\cite{Agashe:2014kda}.
The quantity $\Gamma l(z)$ in function of the redshift has been plotted in Fig.~\ref{param}.

\subsection{Approximate expression for ALP-photon conversions}

\begin{figure}
\centering
\includegraphics[height=12cm, width=12cm]{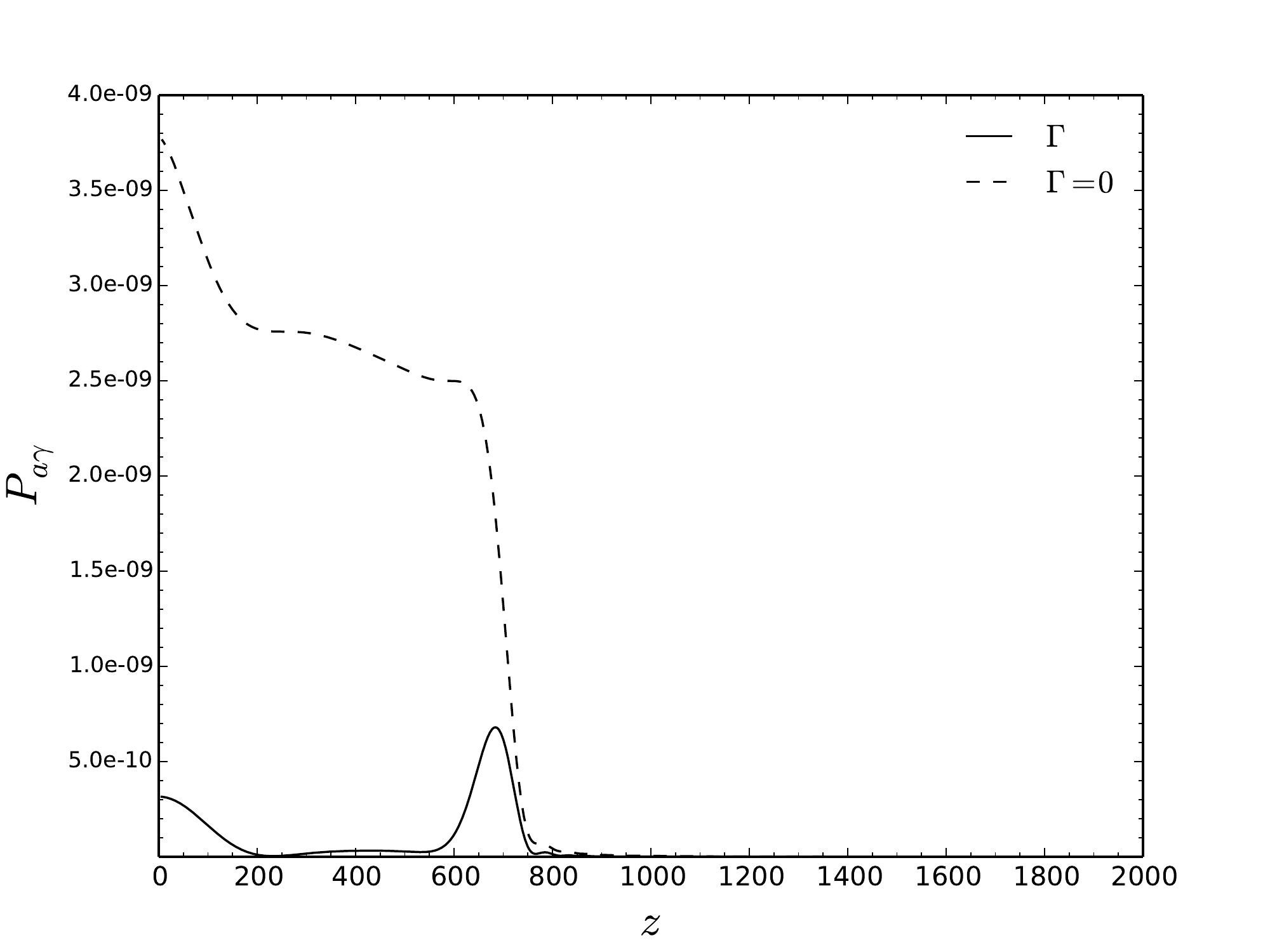}
\caption{Conversion probability $P_{a \gamma}$ in function of the redshift $z$ for ALP energy $E_0=920$~eV
 for $g_{a\gamma} B=10^{-17}  \textrm{GeV}^{-1}  \textrm{nG}$ with absorption $\Gamma$ (continuous curve) and with
 $\Gamma=0$ (dashed curve).
\label{pabs}}
\end{figure}

We finally note that from Fig.~\ref{param} it results that
\begin{equation}
\Delta_{a \gamma} \ll \Delta_{\rm pl} \,\ ,
\end{equation}
independently on the redshift.
This would allow us to find perturbatively a solution of the evolution equation [Eq.~(\ref{vne})] (see Appendix). 
We find a recursive expression for the average conversion probability in the $n$-th magnetic domain
\begin{equation}
P^{(n+1)}_{a\gamma} = \left[P^{(n)}_{a\gamma} + (\Delta^{(n)}_{a \gamma} l_n)^2 
\textrm{sinc}^2\frac{k_n l_n}{2}\right] \text{e}^{-\Gamma_n l_n}
\label{eq:pconvpert}
\end{equation}
where the subscript $n$ refer to the values of the different quantities at redshift $z_n$. 
In the previous equation
\begin{equation}
k=  \Delta_{\rm CMB}+\Delta_{\rm pl}  \,\ ,
\end{equation}
(we remark that $\Delta_{\rm pl}<0$) and the function
\begin{equation}
\textrm{sinc} \,x = \frac{\sin x}{x} \,\ . 
\end{equation}
In Fig.~\ref{pabs} we show the conversion probability $P_{a \gamma}$,  obtained from Eq.~(\ref{eq:pconvpert}),  in function of the redshift $z$ for a representative ALP energy $E_0=920$~eV
 for $g_{a\gamma} B=10^{-17}  \textrm{GeV}^{-1}  \textrm{nG}$ with absorption $\Gamma$ (continuous curve) and with
 $\Gamma=0$ (dashed curve). In the case with $\Gamma=0$ we realize that the conversions in photons occur resonantely at $z \simeq 800$ when $k=0$ (see Fig.~\ref{param}) 
 and then would smoothly grow at a  lower $z$. Instead, in the presence of absorption  the conversions are heavily suppressed just after the resonance point.

\begin{figure}
\centering
\includegraphics[height=12cm, width=12cm]{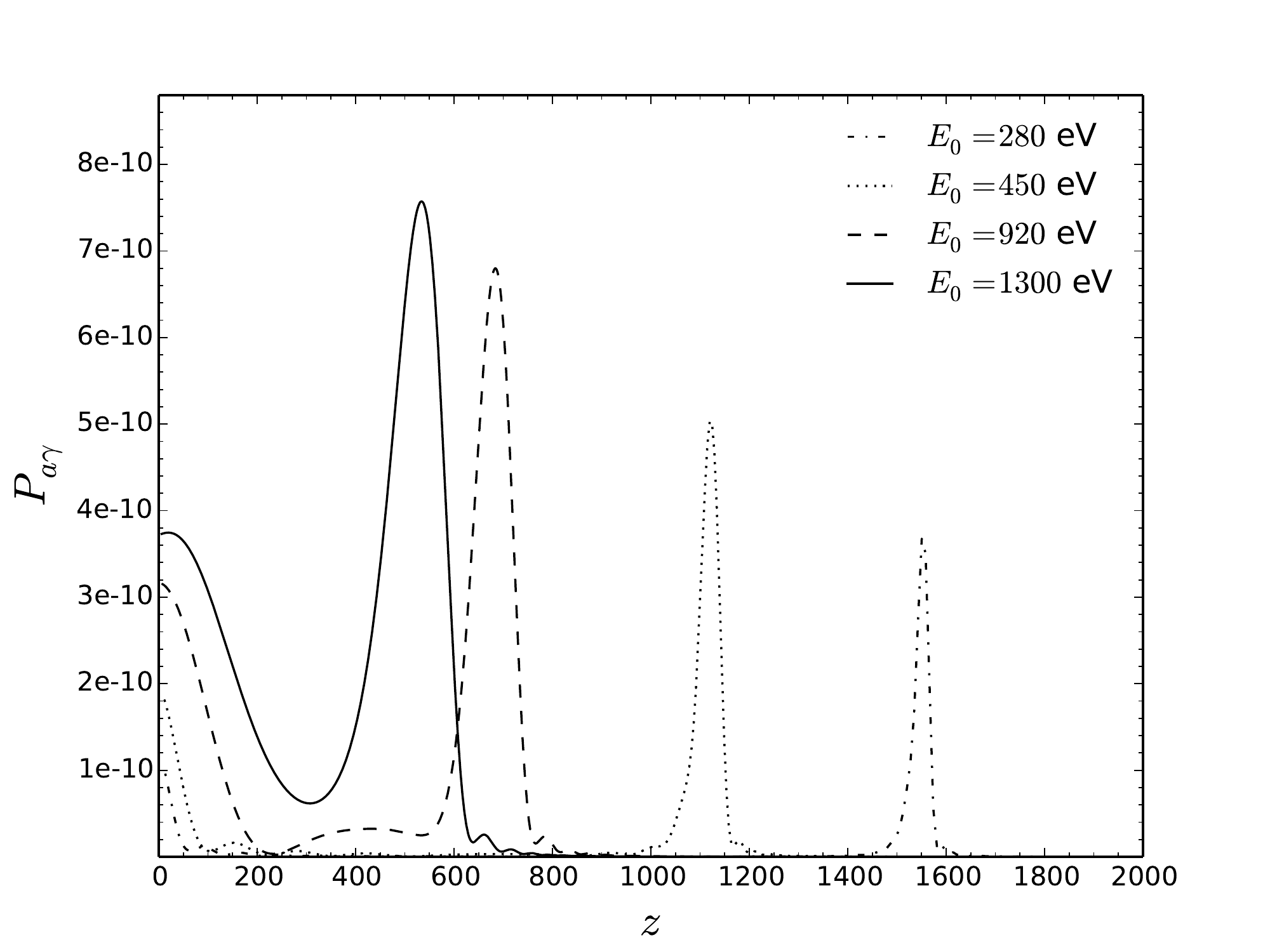}
\caption{Conversion probability $P_{a \gamma}$ in function of the redshift $z$ for different representative energies
of the ALP spectrum for $g_{a\gamma} B=10^{-17}  \textrm{GeV}^{-1}  \textrm{nG}$.
\label{conv}}
\end{figure}

In Fig.~\ref{conv} we show the conversion probability  $P_{a \gamma}$ for 
$g_{a\gamma} B=10^{-17} \textrm{GeV}^{-1}  \textrm{nG}$
 obtained from Eq.~(\ref{eq:pconvpert})
for the same energies considered in Fig.~\ref{param}. We realize that the probability has a typical resonant
behavior, being peaked at redshift for which $k=0$ shown in Fig.~\ref{param}. 
We also note that for the values of the different parameters 
$P_{a \gamma} \lesssim 10^{-9}$.

\section{Our bound on ALPs from reionization}

\subsection{ALP-induced reionization}

The photon spectrum produced by ALP conversions in the $n$-th cell at redshift $z=z_n$
can be written as
\begin{equation}
\frac{d n_{\gamma}}{d E(z)} =\frac{d n_{a}}{d E(z)}\cdot P_{a\gamma}^{(n)} \,\ ,
\end{equation}
where the values of the different quantities characterizing the ALP spectrum  are reported in Sec.~2.
Of these photons the fraction that interacts in the $n$-th cell is $f_S(z)=1-\textrm{exp}(-\Gamma_n  l_n)$. 
Each of these scattered photons together with the photo-electrons gives rise to an electro-magnetic shower
that would ionize the light atoms.  Assuming for simplicity that only H atoms are ionized, one gets 
that the fraction of free electrons produced in the $n$-th cell 
\begin{equation}
\Delta n^{\rm free}_e(z) = \int d E(z) \frac{d n_\gamma}{dE(z)} 
\times f_S(z) \times f_I(z) \times \frac{E(z)}{{13.6} \,\ \textrm{eV}} \,\ ,
\end{equation}
where numerically \cite{Evoli:2012zz}
\begin{equation}
f_I(z)= a \times\left[1-(X_H(z))^b\right]^c \,\ ,
\end{equation}
with $a=0.3846$, $b=0.5420$ and $c=1.1952$.
 The fraction of ionized H at redshift $z=z_n$ can be expressed as
\begin{equation}
X_{\rm H}(z) = X^0_{\rm H}(z) + \frac{\Delta n^{\rm free}_e(z)}{n_{\rm H}} \,\, ,
\end{equation}
where $X^0_{\rm H}(z)$ is the value without extra-ionization.
Note that in principle free electron would recombine with ionized H atoms. However, we estimated
that for the values of the parameters chosen this effect is always negligible.
 
 In Fig.~\ref{XH} we compare $X_{\rm H}$ for the standard case (continuous curve) and in presence of a primordial ALP flux converting into 
 photons with a coupling of $g_{a\gamma} B=10^{-17}\,  \textrm{GeV}^{-1}\,  \textrm{nG}$ (dashed curve) and 
 $g_{a\gamma} B=5 \times 10^{-18}\,  \textrm{GeV}^{-1}\,\textrm{nG}$  (dot-dashed curve).
We assume that ALPs give $\Delta N_{\rm eff}=0.38$.   In the presence of ALPs
 conversions the drop in $X_{\rm H}$ during the recombination epoch is milder leading to a difference of one order of magnitude
 for the  case of the strongest coupling considered.
  
\subsection{Constraints on the ALP conversions from the optical depth}

In order to assess the maximum allowed contribution from cosmic ALPs to reionization we compare the total optical depth in our scenario with the value measured by the Planck collaboration.
The total optical depth encountered by the photons as they travel to us from the surface of last scattering is given by:
 \begin{equation}
 \tau = \int_0^\infty n_e^{\textrm{free}}(z) \sigma_T \left|\frac{dt}{dz}\right| dz\,\, ,
 \end{equation}
where $n_e^{\textrm{free}}(z)=X_H(z)\cdot n_H$ is the fraction of unbound electrons and $\sigma_T$ is the Thomson cross section.
It is then clear that this observable is sensitive to the number of extra ionizations of neutral Hydrogen or Helium atoms at $z \lesssim 1000$ induced by ALP-photon conversions.

Observational evidences (in particular from the spectra of quasars located at redshift $z \sim 6$~\cite{Fan:2005eq}) allows to infer that intergalactic H and He gases are fully ionized below redshift 6, and He is also doubly ionized below redshift $z=3$~\cite{Syphers:2011uw}.  
The resulting free electron population from us to redshift $z=6$ contributes to the total optical depth with about $\tau_{z<6} = 0.038$~\cite{Cirelli:2009bb}.

Planck observations of temperature and polarization anisotropies data provided us with a measurement of the reionization optical depth \cite{Ade:2015xua} 
\begin{equation}
\tau_{\rm obs} = 0.066 \pm 0.016 \,\ , 
\label{eq:tauplanck}
\end{equation}
at 68 $\%$ CL.
Astrophysical sources (as PopIII stars or Quasars) at $z > 6$ can possibly provide enough ionizing photons (either UV or X-rays) to complete the reionization and explain the observed value.
However, the effective contribution of the different kinds of reionization sources and their evolution with redshift is highly uncertain~\cite{Barkana:2000fd,Ferrara:2014sda}.
As in~\cite{Cirelli:2009bb} we bound the ALP contribution by imposing that:
\begin{equation}
\tau_{\rm ALP} < \tau_{\rm obs}^{2\sigma} - \tau_{z<6} \sim 0.044 \,\, ,
\end{equation}
where $\tau_{\rm ALP}$ is the optical depth integrated for $z > 6$ in the presence of ALP-photon conversions and $\tau_{\rm obs}^{2\sigma}$ is the maximum value for $\tau$ allowed (within $2\sigma$) by Planck measurements. 
 Notice that, in doing so, we assume that ALPs are the only source of reionization earlier than $z = 6$. This is unrealistic, since we observe galaxies at redshift higher than $6$. However it provides a conservative bound. Including the contribution of other reionization sources would lead us to stronger constraints.

\begin{figure}
\centering
\includegraphics[height=10cm, width=10cm]{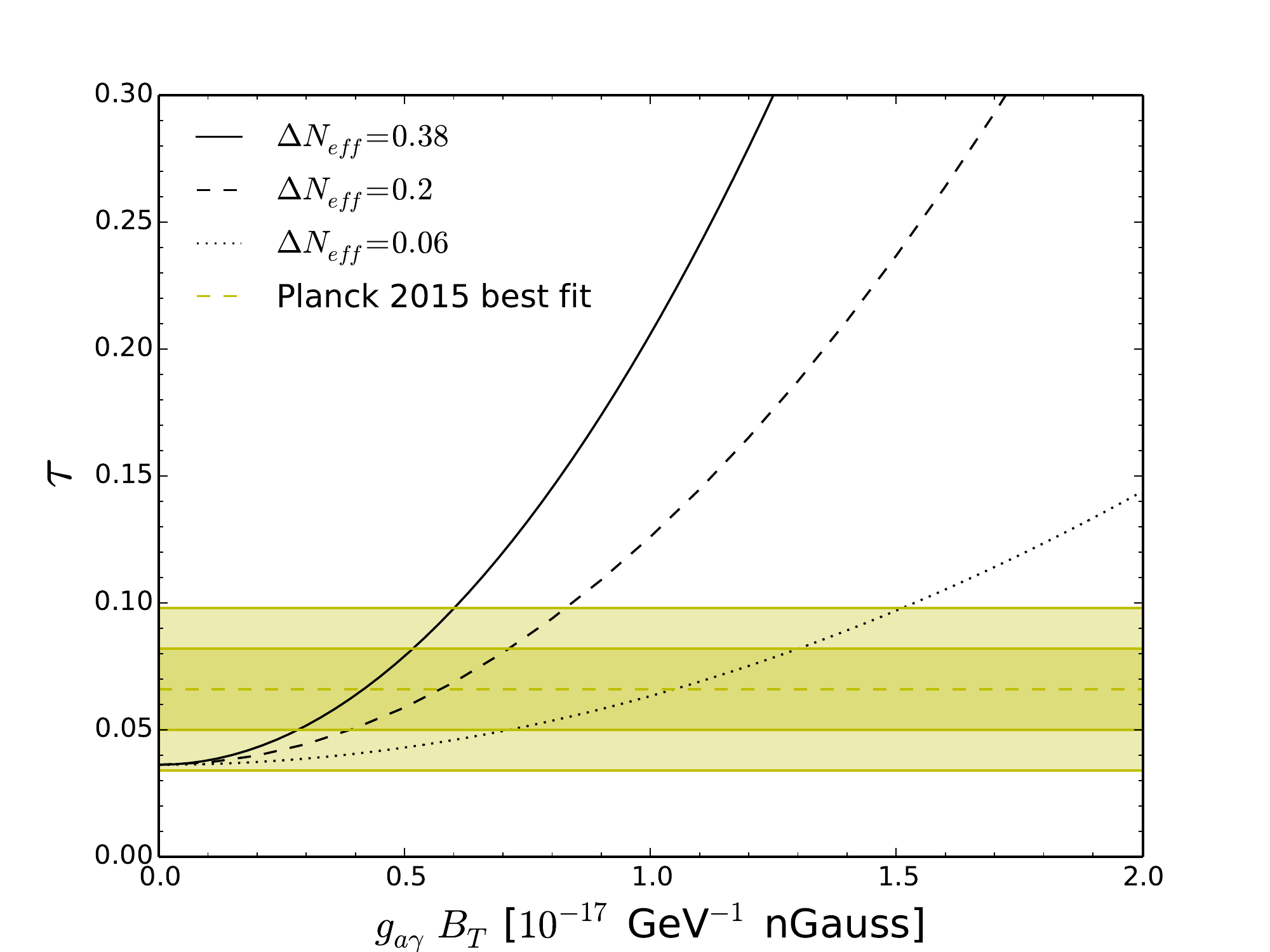}
\caption{Optical depth $\tau$  in presence of ALP for $\Delta N_{\rm eff} = 0.38$ (continuous curve),  $\Delta N_{\rm eff} = 0.2$  (dashed curve) and $\Delta N_{\rm eff}=0.06$ (dotted curve)  in function 
of  $g_{a\gamma} B$. The horizontal bands represent the $1\sigma$ (dark) and $2\sigma$ (light) range for $\tau$ measured from Planck. (see the text for details)
\label{taualp}}
\end{figure}

In Fig.~\ref{taualp} we show the optical depth $\tau$ as a function of  $g_{a\gamma} B$  for 
$\Delta N_{\rm eff} = 0.38$ (continuous curve) in agreement with the $1\sigma$ range from Planck 2015 analysis  \cite{Ade:2015xua},
  $\Delta N_{\rm eff} = 0.2$  (dashed curve) and $\Delta N_{\rm eff}=0.06$ (dotted curve)  which corresponds to the $2\sigma$ forecast of the future EUCLID experiment \cite{Basse:2013zua}.
Horizontally we show the $2\sigma$ band of $\tau$ from Eq.~(\ref{eq:tauplanck}).
We realize that  for $\Delta N_{\rm eff} = 0.38$ values of   $g_{a\gamma} B \gtrsim 6 \times 10^{-18}\,  \textrm{GeV}^{-1}\,  \textrm{nG}$
would be excluded. The bound is worsened if less amount of extra-radiation is composed by ALPs going down to 
$g_{a\gamma} B \gtrsim 1.5 \times 10^{-17}\,  \textrm{GeV}^{-1}\,  \textrm{nG}$ for  $\Delta N_{\rm eff}=0.06$.
Therefore, also in the presence of a subleading content of ALP extra-radiation the bound remains impressive.

\section{Discussion and Conclusions}

String theory would provide an intriguing connection between the possible extra-radiation $\Delta N_{\rm eff}$ and
a flux of ultralight ALPs generated by heavy moduli decays in the post-inflation epoch.
The presence of this cosmic background would have an interesting phenomenology. Indeed
 X-ray excesses in many Galaxy Clusters may be explained by the conversions of the
this cosmic ALP background radiation into photons in the Cluster magnetic field.
In the current work we have explored another possible feature of this model.
Indeed, the presence of primordial magnetic fields in the Early Universe would trigger ALP-photon conversions during the dark ages,
producing a flux of sub-MeV photons that would ionize the recently formed H atoms. Impressively also conversions 
at level 
$P_{a\gamma} \lesssim 10^{-9}$  would be enough to produce a seizable effect. 
Comparing this effect withe the measurement of the optical depth in the Universe by the Planck experiment, we have shown 
that in principle it would be possible to obtain strong bounds on the ALP-photon coupling 
\begin{equation}
g_{a\gamma} B \lesssim 6 \times 10^{-18}\,  \textrm{GeV}^{-1}\,  \textrm{nG} \,\ ,
\label{eq:bound}
\end{equation}
assuming that ALPs compose all the extra-radiation compatible with the $1\sigma$ bound of Planck
($\Delta N_{\rm eff}=0.38$). Intriguingly, if the primordial magnetic field would have a value close to the present limits ($\sim$nG), Eq.~(\ref{eq:bound}) sets a limit on $g_{a\gamma}$ close to inverse of the Planck scale.
(At this scale  it has shown in  \cite{Raffelt:1987im}  that the graviton would take the role of an ALP with $g_{a\gamma}\sim M_{\rm pl}^{-1}$).

We stress that without direct evidence for a primordial magnetic field, our bounds do not allow to constrain directly the coupling constant $g_{a \gamma}$. 
However, even if a primordial magnetic field would be measured with values much lower than the current upper bound, $B \sim$~nG, 
Eq.~(\ref{eq:bound}) would strongly constrain the phenomenological consequences of the  cosmic ALP background. 
 Indeed, the X-ray excess would typically require a coupling $g_{a\gamma} \sim 10^{-13} \, \textrm{GeV}^{-1}$ \cite{Conlon:2013txa,Angus:2013sua,Powell:2014mda}.  Conversely, if the ALP interpretation of X-ray excess 
 would be confirmed, 
Eq.~(\ref{eq:bound})  would provide the  strongest constraint of the primordial magnetic field at the recombination epoch.
 Therefore, it is nice to realize that our cosmological limits from reionization could have relevant consequences for signatures of  cosmic 
ALPs in X-ray sources. This confirms once more the nice synergy between astrophysics and cosmology to search axion-like particles.

\section*{Acknowledgments}

We acknowledge Andrea Ferrara, Tetsutaro Higaki, Kazunori Nakayama, Fuminobu Takahashi and David J.\ E.\ Marsh for useful comments on the manuscript.
The work of A.M.~and D.M.~is supported by the Italian Ministero dell'Istruzione, Universit\`a e Ricerca (MIUR) and Istituto Nazionale di Fisica Nucleare (INFN) through the ``Theoretical Astroparticle Physics'' project. The work of M.L. is supported by the European Research Council under ERC Grant ``NuMass'' (FP7- IDEAS-ERC ERC-CG 617143).

\section*{Appendix: Approximate expression for ALP-photon conversion probability}

In this Appendix we present the derivation of an approximate expression for the ALP-photon conversion probabilities
suitable for our purposes. 
We realize that for the input parameters chosen in Eq.~(\ref{aa8MR})
it results
\begin{equation}
\Delta_{a\gamma}\ll \Delta_{\rm pl}.\label{eq:limitedeltaosc}
\end{equation}
This hierarchy is valid independently of the redshift since 
$\Delta_{a\gamma}$
and  $\Delta_{\rm pl}$ both scale as
$\left(1+z\right)^{2}$ (see Fig.~\ref{param}).
Including  the 
absorption term $\Gamma$ and neglecting $\Delta_{\rm QED}$, 
we rewrite the Hamiltonian [Eq. (\ref{eq:massgen})],  up to an overall
term proportional to $\Delta_{a}\mathbb{I}_{3\times3}$ as 
\begin{equation}
{\mathcal M}_p=\left(\begin{array}{ccc}
k-i\frac{\Gamma}{2} & 0 & \Delta_{a\gamma}\cos\psi_p\\
0 & k-i\frac{\Gamma}{2} & \Delta_{a\gamma}\sin\psi_p\\
\Delta_{a\gamma}\cos\psi_p & \Delta_{a\gamma}\sin\psi_p & 0
\end{array}\right),
\end{equation}
where we have defined  $k\equiv\Delta_{\rm CMB}+\Delta_{\rm pl} -\Delta_{a}$.
Introducing this change of variables
\begin{equation}
A_{1,2}=\hat{A}_{1,2}\exp\left\{ -i\int_{0}^{x_{3}}ds\left(k(s)-i\frac{\Gamma(s)}{2}\right)\right\} \equiv\hat{A}_{1,2}\text{e}^{-\alpha(x_{3})},
\end{equation}
the evolution equations  acquire the form
\begin{equation}
i\frac{\partial}{\partial x_{3}}\left(\begin{array}{c}
\hat{A}_{1}\\
\hat{A}_{2}\\
a
\end{array}\right)=\left(\begin{array}{ccc}
0 & 0 & \Delta_{a\gamma}\cos\psi_p\, \text{e}^{\alpha}\\
0 & 0 & \Delta_{a\gamma}\sin\psi_p\, \text{e}^{\alpha}\\
\Delta_{a\gamma}\cos\psi_p\, \text{e}^{-\alpha} & \Delta_{a\gamma}\sin\psi_p\, \text{e}^{-\alpha} & 0
\end{array}\right)\left(\begin{array}{c}
\hat{A}_{1}\\
\hat{A}_{2}\\
a
\end{array}\right).\label{eq:evolution-2-1-1}
\end{equation}
Since $\Delta_{a\gamma}\ll\Delta_{pl}$ and $\Delta_{a\gamma}x_{3}\ll1$
we can expand at the first order the solution of the system  (\ref{eq:evolution-2-1-1})
\begin{eqnarray}
\hat{A}_{1}(x_{3}) &=&-i\int_{0}^{x_{3}}d\eta\Delta_{a\gamma}\cos\psi_p(\eta)\, \text{e}^{\alpha(\eta)}a(0) \nonumber \\
\hat{A}_{2}(x_{3}) &=& -i\int_{0}^{x_{3}}d\eta\Delta_{a\gamma}\sin\psi_p(\eta)\, \text{e}^{\alpha(\eta)}a(0) \,\ .
\label{eq:hatA}
\end{eqnarray}
We assume that initially the beams is composed by only ALPs, so that $a(0)=1$, while $\hat{A}_{1,2}(0)=0$. 
Then the probability that an ALP is converted into a photon is given by
\begin{equation}
P_{a\rightarrow\gamma}(x_{3})=\left|A_{1}(x_{3})\right|^{2}+\left|A_{2}(x_{3})\right|^{2}=\left( \left|\hat{A}_{1}(x_{3})\right|^{2}+\left|\hat{A}_{2}(x_{3})\right|^{2}\right) \text{e}^{-\int_{0}^{x_{3}}ds\Gamma(s)} \,\ ,
\end{equation}
where from Eq. (\ref{eq:hatA})
\begin{equation}
\hat{A}_{1}(x_{3})=-i\frac{g_{a\gamma}}{2}\int_{0}^{x_{3}}d\eta B_{1}(\eta)\, \text{e}^{\alpha(\eta)},\label{eq:hatA-1}
\end{equation}
with $B_{1}(\eta)=B_T(\eta)\cos\psi_p(\eta)$. 
Assuming a cell-like structure for the $B$-field, the previous integral can be written as
\begin{equation}
\hat{A}_{1}(x_{3})=-i\frac{g_{a\gamma}}{2}\sum_{p=1}^{n}B_{1,p}\int_{x_{3,p}}^{x_{3,p+1}}d\eta\, \text{e}^{\alpha(\eta)},\label{eq:hatA-1-1}
\end{equation}
where $[x_{3,p},x_{3,p+1}]$ is the interval of  $p$-th and $B_{1,p}$ is the value of the magnetic field therein.
Then, averaging over all the possible magnetic field configuration we get
\begin{equation}
\left\langle \left|\hat{A}_{1}(x_{3})\right|^{2}\right\rangle =\frac{g_{a\gamma}^{2}}{8}\sum_{p=1}^{n}(B^2_T)_p \int_{x_{3,p}}^{x_{3,p+1}}d\eta_{1}d\eta_{2}\, \text{e}^{i\int_{\eta_{1}}^{\eta_{2}}ds \, k(s)}\text{e}^{\int_{0}^{x_{3,p}}ds\, \Gamma(s)},
\end{equation}
where we used that $\alpha\equiv k-i\Gamma/2$ and assumed $\Gamma(s)$
constant in each cell.
Assuming also 
$k(s)$ constant in each cell, we get
\begin{equation}
\left\langle \left|\hat{A}_{1}(x_{3})\right|^{2}\right\rangle =\frac{g_{a\gamma}^{2}}{8}\sum_{p=1}^{n}(B^2_T)_p \int_{x_{3,p}}^{x_{3,p+1}}d\eta_{1}d\eta_{2}\, \text{e}^{ik(\eta_{1})\left(\eta_{2}-\eta_{1}\right)}\text{e}^{\int_{0}^{x_{3,p}}ds \,\Gamma(s)}.
\end{equation}
An analogous expression occurs for $\left\langle \left|\hat{A}_{2}(x_{3})\right|^{2}\right\rangle$.
Then the conversion probability into photons assumes the form
\begin{equation}
P_{a\rightarrow\gamma}(x_{3})=\frac{g_{a\gamma}^{2}}{4}\sum_{p=1}^{n}(B^2_T)_p \left|\int_{x_{3,p}}^{x_{3,p+1}}d\eta\, \text{e}^{-ik(\eta)\eta}\right|^{2}\text{e}^{-\int_{x_{3,p}}^{x_{3}}ds \, \Gamma(s)} \,\ .
\end{equation}
Using the Fourier transform of the hat function, this equation can be written as
\begin{equation}
P_{a\rightarrow\gamma}(x_{3})=\frac{g_{a\gamma}^{2}}{4}\sum_{p=1}^{n} (B^2_T)_p \,l_{p}^{2}\text{ \text{sinc}}^{2}\left(\frac{k_{p}l_{p}}{2}\right)\text{e}^{-\int_{x_{3,p}}^{x_{3}}ds\, \Gamma(s)},\label{eq:probintere}
\end{equation}
where $\text{sinc}(x)\equiv\sin(x)/x$,
$l_{p}=\left|x_{3,p+1}-x_{3,p}\right|$ is the size of each  cell and
$k_{p}\equiv k(x_{3,p})$. 
Finally, we can write a recursive expression for the conversion probability in the $n$-th magnetic domain
\begin{equation}
P^{(n+1)}_{a\gamma} = \left[P^{(n)}_{a\gamma} + (\Delta^{(n)}_{a \gamma} l_n)^2 
\textrm{sinc}^2\frac{k_n l_n}{2}\right] \text{e}^{-\Gamma_n l_n}
\end{equation}
where the subscript $n$ refer to the values of the different quantities at redshift $z_n$. 

A brief remark is in order.  A very similar expansion can be obtained in the case of a more realistic homogeneous and isotropic turbulent magnetic field, where the Fourier components are correlated as~\cite{Mirizzi:2007hr}
\begin{equation}
\langle{\tilde B}_i({\mathbf k}){\tilde B}_j({\mathbf k}')\rangle=(2\pi)^6 M(|{\mathbf k}|)\cdot\left(\delta_{ij}-\frac{k_i k_j}{|{\mathbf k}|^2}\right)\delta^3({\mathbf k}-{\mathbf k}')\, ,
\end{equation}
(where the tensor in bracket implements the condition $\nabla\cdot\mathbf{B}=0$). At first order, after straightforward calculations we obtain
\begin{equation}
P_{a\rightarrow\gamma}(x_{3})=\frac{g_{a\gamma}^{2}}{2}\int_{0}^{x_{3}}ds\,\tilde{\varepsilon}_{\perp}(k(s))\,\text{e}^{-\int_{s}^{x_{3}}du\,\Gamma(u)}\label{eq:Probevol1}\, ,
\end{equation}
(we omit the proof), where $\tilde{\varepsilon}_{\perp}(k)$ is the Fourier transform of the correlation function ${\mathcal C}(y)=\langle B_1(x_3) B_1(x_3+y)\rangle$ along the line-of-sight. 

For a cell-like structure $\langle B_1(x_3) B_1(x_3+y)\rangle=\langle B_{1,p}(x_3) B_{1,p}(x_3+y)\rangle$ where $p$ is any of the cells along the line-of-sight, while the correlation between adjacent cells is zero. Since $B_{1,p}(x_3)$ is a hat function with width $l$ we have ${\mathcal C}(y)=B_T^2 \cdot (1-|y|/l)$ for $|y|\leq l$ and zero otherwise, whose Fourier transform is just 
\begin{equation}
\tilde{\varepsilon}_{\perp}(k)=\frac{1}{2}B_T^2\, l\,\textrm{sinc}^2\left(\frac{k l}{2}\right)\, .
\end{equation}
In this case Eq.~(\ref{eq:Probevol1}) returns Eq.~(\ref{eq:probintere}). For a Kolmogorov-like power-law spectrum $M(|{\mathbf k}|)\sim |{\mathbf k}|^q$ with $k_L\leq  |{\mathbf k}|\leq k_H$ the function $\tilde{\varepsilon}_{\perp}(k)$ has been calculated in~\cite{Meyer:2014epa} [Eqs.~(A.9--16)].
We have explicitly checked that the results are similar to those obtained for a cell-like structure for several choices of the spectral index $q$ and cut-off parameters $k_{H,L}$. This shows that the choice of the cell-like structure for the magnetic field is not crucial for our results.

\end{document}